\documentclass[aps,twocolumn,prb, groupedaddress]{revtex4-1}

\usepackage{graphicx}
\usepackage{graphics}
\usepackage{amssymb}
\begin{document}

\title{Direct observation of the superconducting gap in thin film of titanium nitride using terahertz spectroscopy}

\author{Uwe S. Pracht}
\email[electronic adress ]{uwe.pracht@pi1.physik.uni-stuttgart.de}
\affiliation{1. Physikalisches Institut, University of Stuttgart, Pfaffenwaldring 57, 70550 Stuttgart, Germany}
\author{Marc Scheffler}
\affiliation{1. Physikalisches Institut, University of Stuttgart, Pfaffenwaldring 57, 70550 Stuttgart, Germany}
\author{Martin Dressel}
\affiliation{1. Physikalisches Institut, University of Stuttgart, Pfaffenwaldring 57, 70550 Stuttgart, Germany}
\author{David F. Kalok}
\affiliation{ Institute of Experimental and Applied Physics, University of Regensburg, 93025 Regensburg, Germany} 
\author{Christoph Strunk}
\affiliation{ Institute of Experimental and Applied Physics, University of Regensburg, 93025 Regensburg, Germany} 
\author{Tatyana I. Baturina}
\affiliation{A. V. Rzhanov Institute of Semiconductor Physics SB RAS, 13 Lavrentjev Avenue, Novosibirsk 630090, Russia}

\date{\today}

\begin{abstract}
We report on the charge carrier dynamics of superconducting titanium nitride (TiN) in the frequency range  90 - 510 GHz (3 - 17 cm$^{-1}$).  The experiments were perfomed on a 18 nm thick TiN film with a critical temperature of $T_c=3.4$ K. Measurements were carried out from room temperature down to 2 K, and in magnetic fields up to $B=7$ T. We extract the real and imaginary parts of the complex conductivity $\hat{\sigma}$ as a function of frequency and temperature, directly providing the superconducting energy gap $2\Delta$. Further analysis yields the superconducting London penetration depth $\lambda_L$. The findings as well as the normal state properties strongly suggest conventional BCS superconductivity, underlined by the ratio $2\Delta(0)/k_BT_c=3.44$. Detailed analysis of the charge carrier dynamics of the silicon substrate is also discussed.
\end{abstract}

\maketitle

\begin{figure}
\includegraphics[width=8.66cm]{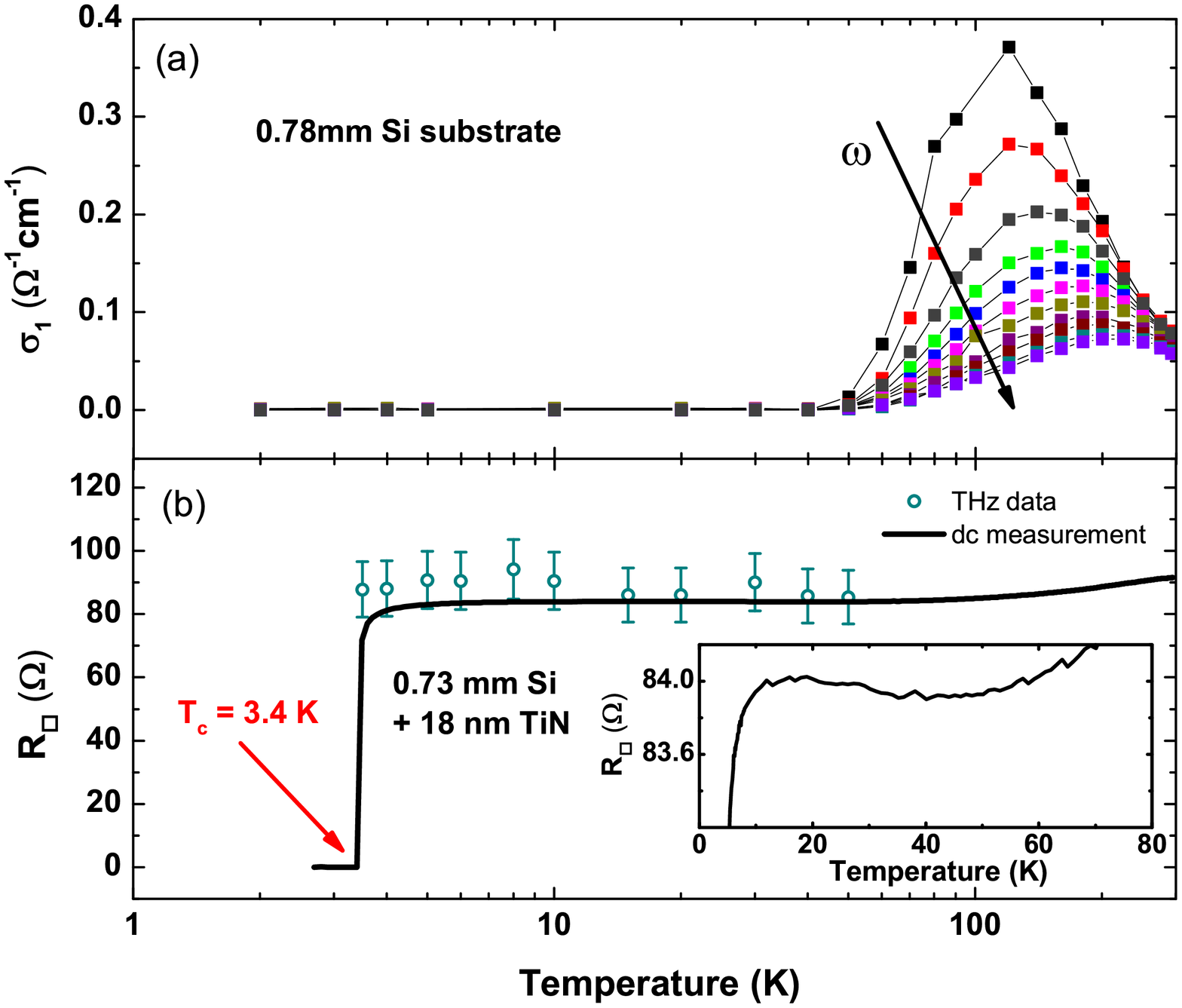}
\caption {\label{fig1}(color online) (a) Real part of the substrate conductivity versus temperature for different frequencies from  4-23 cm$^{-1}$. Below 50 K, the conductivity drops to zero and the substrate remains transparent (b) Sheet resistance of the TiN film versus temperature. We emphasize the good agreement between dc and THz approach (solid line and open circles respectively). The Inset shows a close up of the sheet resistance. The small peak is a signature of quantum contributions.}
\end{figure}

For the last few years promising research has been done on highly yet homogenously disordered TiN superconducting thin films, since this material has been identified to be a prime candidate for the superconductor-insulator transition\cite{Gan10} (SIT). Besides fundamental physical interest, TiN has recently gained attention as a material for microwave resonators\cite{vis12}. Today, one can access a broad range of experimental data displaying the dc-properties of TiN superconducting thin films, e.g. comprehensive dc-transport experiments or local scanning tunneling spectroscopy at millikelvin temperatures and high magnetic fields. Very intriguing phenomena like a energy gap\cite{sac08} for $T>T_c$ which might serve as key for understanding the pseudogap\cite{bat10} in high-$T_c$ superconductors, voltage threshold behaviour\cite{bat07, bat07b} or a notable peak in the magnetoresistance\cite{bat05,bat07b} have been observed at the verge of the SIT. All these phenomena, as well as the very nature of the SIT itself, are not yet entirely understood. For the last years, several theoretical approaches were studied\cite{fei10}. One common belief is that superconductivity is destroyed on a macroscopic level when entering the normal state. However, microscopic islands embedded into an insulating or metallic matrix remain superconducting but have lost any mutual phase correlation. This picture is supported experimentally by magnetoresistance effects\cite{dub05}, local STM\cite{sac08}, voltage threshold behaviour\cite{bat07, bat07b} and numerically by Monte-Carlo simulations\cite{bou11}.
There is, however, very little known about the electrodynamic response of superconducting thin films, which are known to show the SIT. Infrared (IR) reflectivity measurements\cite{pfu09} were performed on several TiN thin film samples with different degrees of disorder and broadwave microwave corbino spectroscopy on various InO thin film samples\cite{Arm11}. With our work, we extend the studied frequency range well below the far IR down to 3 $\text{cm}^{-1}$. This range provides direct access to the superconducting properties, e.g. the real and imaginary parts of the complex conductivity, the energy gap and the London penetration depth as functions of frequency and temperature below and above the energy gap.
Identical experiments were performed with two similar TiN films. Since both TiN films lead to the same results, we will just discuss one of them. The presented data was taken on a $5 \times 5$ mm$^2$ large TiN film of 18 nm thickness grown via plasma enhanced atomic layer deposition  at $T=400$ $^{\circ}$C on a 0.73 mm thick silicon (110) substrate covered with a 10 nm thick SiO$_2$ film.  The critical temperature is $T_c=3.4$ K and the sheet resistance  $R_{\square}(300\text{K})=91.6$ $\Omega$. Considering these values, we expect a TiN film, which is at the boundary between bulk and quasi-two-dimensional superconductivity. Fig. \ref{fig1}b displays the sheet resistance $R_{\square}$ of the TiN film as a function of temperature. The data was taken in a standard 4-point-measurement using Van-der-Pauw analysis\cite{vdp58}. Starting at room temperature, $R_{\square}$ decreases linearly from 91.6 $\Omega$ to 85 $\Omega$. Immediately before the superconducting transition, a small peak shows up at 15 K, see the inset in Fig. 1b. This peak emerges from the competition between weak localization, electron-electron interaction and supercondcuting fluctuations. In TiN films with a higher degree of disorder this peak is much more pronounced\cite{bat12} while it is absent in clean films. Since the absolute value of quantum contributions to conductivity is much smaller than conductivity itself in what follows we will not take it into account.  At 4 K, $R_{\square}$ decreases rapidly and drops to immeasurably small values at $T_c=3.4$ K. 

We performed transmission and phaseshift measurements covering the spectral range  3 - 17 $\text{cm}^{-1}$ for 2-292 K and magnetic fields up to 7 T. Tuneable backward-wave-oscillators (BWO) emitted coherent radiation of high power, which was then detected by a Golay cell. The TiN film sample and a bare Si substrate sample as reference were mounted into an optical $^4$He bath cryostat oriented as shown in the inset of Fig. 2a.  Both transmission and phaseshift were measured with the same quasioptical Mach-Zehnder interferometer\cite{dre08, Ost11, tor12}. Transmission $T_f$ and phaseshift over frequency $\phi_t/\omega$ were fitted simultaneously using well known equations based on Fresnel´s formulas for multiple reflections\cite{dre02}. Fitting parameters were the real and imaginary parts of the refractive index, $n$ and $k$, respectively. Considering these two quantities, one obtains any optical function desired, e.g. the complex conductivity $\hat{\sigma}$ as follows
\begin{equation}
\sqrt{n+ik}=\epsilon_1(\omega)+i\epsilon_2(\omega)=1-\frac{4\pi}{\omega}\sigma_2(\omega)+\frac{4\pi i}{\omega}\sigma_1(\omega)
\end{equation}
The open circles in Fig. \ref{fig1}b is the sheet resistance\cite{Kit95} $R_{\square}= (d\sigma_0)^{-1}$ ($d=18$ nm film thickness) extracted from the THz data. We obtain the same results for $R_{\square}$ within an accuracy of a few percent for both dc and THz approach. Fig. 1a shows real parts of the complex conductivity of the silicon substrate versus temperature for several frequencies between 4-23 cm$^{-1}$. Below 50 K, $\sigma_1$ drops to zero and remains there down to our base temperature of 2 K. 
This implies that the substrate is purely dielectric in the superconducting regime of TiN. A detailed discussion of the THz properties of the substrate is provided in the appendix.
Fig. 2 displays $T_f$ and $\phi_t/\omega$ versus frequency for 2 K in the superconducting and 10 K in the normal state. The solid line is a calculation according to the BCS theory applied to both $T_f$ and $\phi_t/\omega$ simultaneously, with $n$ and $k$ expressed by $\sigma_1$ and $\sigma_2$  of the Mattis-Bardeen formalism\cite{mat58, Zim91} (dirty limit). The oscillations are caused by interference due to multiple reflections inside the substrate, which behaves like a nonsymmetric Fabry-Perot resonator. 
\begin{figure}
\includegraphics[width=8.66cm]{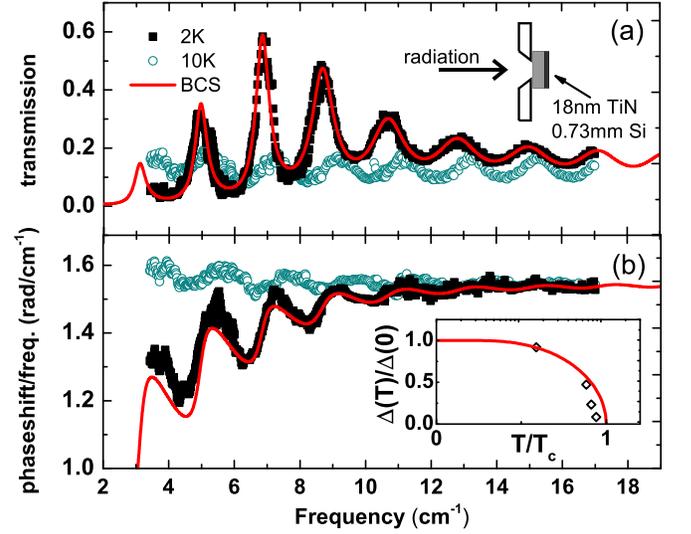}
\caption {\label{fig2}(color online) (a) Transmission $T_f$  and (b) phaseshift over frequency $\phi_t/\omega$ versus frequency at 2 K and 10 K below and above $T_c=3.4$ K. The solid line is a calculation according to BCS theory. The change in both properties indicate a strong frequency dependent complex conductivity. The inset shows the temperature evolution of the reduced energy gap together with the BCS prediction leading to $2\Delta(0)=8.14$ cm$^{-1}$.}
\end{figure}
At 10 K, the Fabry-Perot oscillations show a uniform pattern, which does not vary with frequency. As implied by the sheet resistance in Fig. 1b, this is due to the dispersionless metallic conductivity of the TiN film at frequencies much lower than the scattering rate. As soon as the critical temperature is passed, $T_f$ and $\phi_t/\omega$ change dramatically, and $T_f$ rises well above the normal conducting values at 2 K and for frequencies larger than ~6 cm$^{-1}$. This can be understood in the two-fluid model: the number of quasi particles contributing to $\sigma_1$ is reduced due to the formation of Cooper pairs, which do not contribute to $\sigma_1$ for finite frequencies below $2\Delta$. For frequencies lower than ~6 cm$^{-1}$, $T_f$ drops below the normal state $T_f$ because of the increasing reflection since the superconducting film acts like a dielectric mirror. $\phi_t/\omega$, however, decreases for all frequencies. The considerable changes below $T_c$ allude a strong frequency dependent complex conductivity. 

From the raw data we extracted the superconducting properties of the TiN film: we perform single-peak fits for each Fabry-Perot oscillation separately. For each peak, i.e. each resonance frequency, we obtain a pair of $\epsilon_1$ and $\epsilon_2$, which is then used to calculate the complex conductivity. 
\begin{figure}
\includegraphics[width=8.66cm]{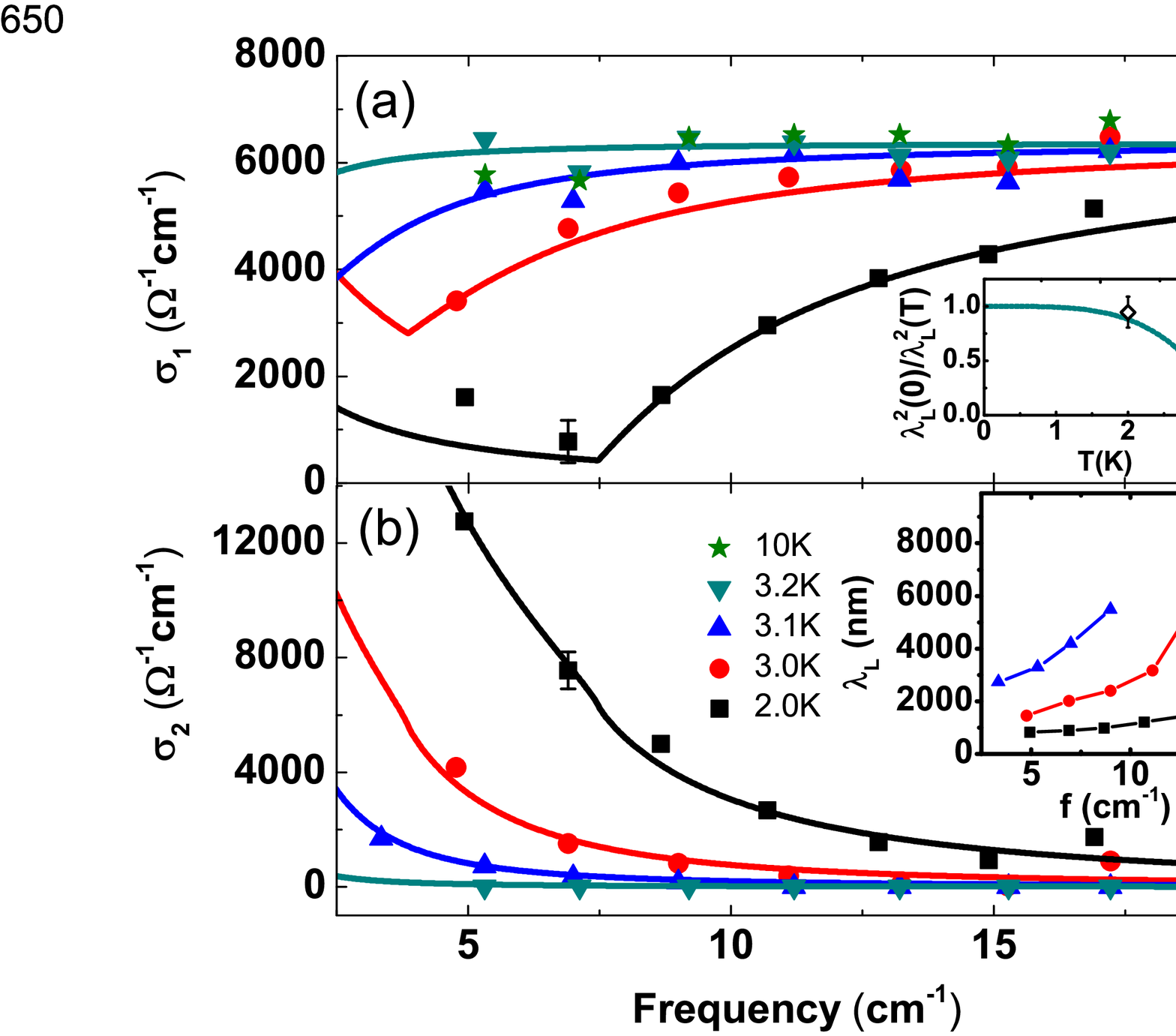}
\caption {\label{fig3}(color online) (a) Real and (b) imaginary parts of the complex conductivity versus frequency for four different temperatures below $T_c=3.4$ K and one well above. The solid  lines are fits according to BCS theory applied to both $\sigma_1$ and $\sigma_2$ simultaneously. The kink in $\sigma_1$ indicates the energy gap, which shifts towards lower frequencies upon increasing temperature. The insets show the London penetration depth versus frequency and temperature, respectively. The solid line in the upper inset is the two-fluid model prediction yielding $\lambda_L(0)=730$ nm.} 
\end{figure}
Real and imaginary parts of $\hat{\sigma}(\omega)$ are shown in Fig. 3 for several temperatures below and above $T_c$. The solid lines are fits according to Mattis-Bardeen theory\cite{mat58, Zim91} (dirty limit) applied to both $\sigma_1$ and $\sigma_2$ simultaneously. As can be seen, $\sigma_1$ and $\sigma_2$ are well described by the BCS theory. Starting from 2 K, $\sigma_1$ is strongly suppressed for frequencies around 7 cm$^{-1}$ and approaches the normal state frequency independent conductivity $\sigma_0=6350\pm400$ $\Omega^{-1}\text{cm}^{-1}$. The minimum in $\sigma_1$ indicates to the energy gap that gets less pronounced and shifted towards lower frequencies upon increasing temperature. This trend is displayed in the inset of Fig. 2b together with the BCS expression\cite{Ric65}
\begin{equation}
\frac {\Delta(T)}{\Delta(0)}=\tanh\left\{\frac{T_c}{T}\frac{\Delta(T)}{\Delta(0)}\right\}
\end{equation}
The fit leads to a zero temperature gap of $2\Delta(0)=8.14$ $\text {cm}^{-1}$ and consequently a BCS ratio of  $2\Delta(0)/k_BT_c=3.44$ very close to the weak-coupling prediction\cite{tin80} of 3.53. Upon cooling, $\sigma_2$ rises notably from zero and approaches $\sigma_2\propto \omega ^{-1}$, which is consistent with   $\sigma_1 \propto \delta(\omega=0)$  at zero temperature according to Kramers-Kronig relations. This trend is also seen in the frequency dependence of the London penetration depth $\lambda_L$, which was calculated from $\sigma_2$  via\cite{dre02}
\begin{equation}
\lambda_L=\sqrt \frac{c^2}{4\pi \omega \sigma_2}
\end{equation}
Assuming the above mentioned $\omega^{-1}$-behavior, $\lambda_L$ gets frequency independent. In contrast to $T\lesssim T_c$, one finds a nearly constant $\lambda_L$ at 2 K, which underlines the theoretical prediction. For $T\lesssim T_c$, however, $\lambda_L$ shows a strong rise towards high frequencies. We attribute these high penetration depths to the increasing number of quasiparticles due to thermal and photon induced Cooper-pair breaking. Upon increasing photon energy $\hbar \omega>2\Delta$,  Cooper pairs are destroyed, and hence the superconducting state is weakend. Extrapolating the $\lambda_L(\omega)$ curves to zero frequency leads to the data shown in the inset of Fig. 3a, where the squared ratio $\lambda_L(0)/\lambda_L(T)$ versus temperature is shown. The data was fitted within the two-fluid model\cite{tin80}
\begin{equation}
\frac{\lambda_{L}^{2}(0)}{\lambda_{L}^{2}(T)}=1-\left(\frac{T}{T_{c}}\right)^{4}
\end{equation}
The fit yields the London penetration depth $\lambda_L(0)=730\pm50$ nm at zero temperature. 
\begin{figure}
\includegraphics[width=8.66cm]{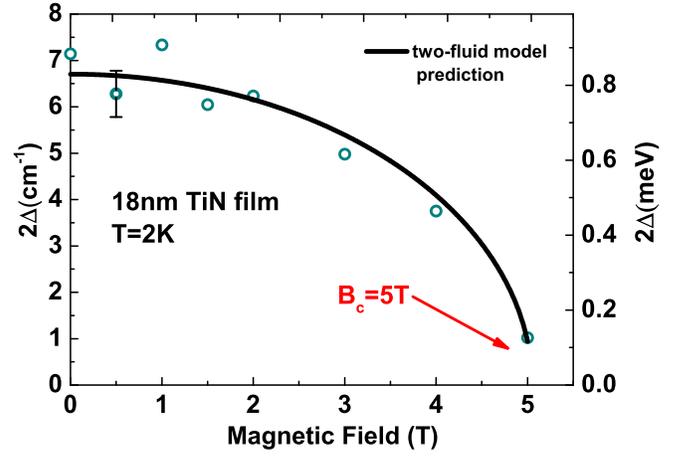}
\caption {\label{fig4}(color online) Energy gap versus magnetic field (open circles) together with the prediction based on the phenomenological two-fluid model. The fit leads to  $B_c=5$ T. }
\end{figure}

The data presented so far strongly supports the interpretation as a conventional BCS superconductor. We also performed experiments in magnetic fields up to $B=7$ T at 2 K utilizing a superconducting cryomagnet optical access and studied the influence on the energy gap. The magnetic field was applied parallel to the film. The energy gap was extracted by fitting $T_f$ and $\phi_t/\omega$ simultaneously for each magnetic field. The fitting parameters were $2\Delta$ and $T_c$ while the other material parameters, plasma frequency $\omega_P$ and scattering rate $\Gamma$, utilzed for the fit were calculated separately from the normal state at 4 K and were taken as constants.  Fig. 4 displays energy gap versus magnetic field together with a fit via the phenomenological approach\cite{tin80}
\begin{equation}
\frac{\Delta(B)}{\Delta(0)}=\sqrt{1-\left(\frac{B}{B_{c}}\right)^{2}}
\end{equation}
yielding a (upper) critical field of $B_c=5$ T. The gap decreases monotonically with increasing magnetic field until the pair-breaking limit is reached and superconductivity is totally suppressed. The analysis of the raw data for higher fields did not indicate an energy gap. The zero-field value of $2\Delta$ of the different setups are in good agreement (7.5 cm$^{-1}$  and 7.2 cm$^{-1}$). 

In summary, we have studied the charge carrier dynamics of superconducting TiN films with a low degree of disorder at temperatures well below $T_c=3.4$ K and frequencies above and below $2\Delta$. The superconducting properties are well described within the BCS theory. Future efforts will have to extend the experimentally accessible range towards lower temperatures to shed light on the yet not understood phenomena close to the SIT. 

We would like to thank K. Sedlmeier and B. Gorshunov for help with the experiments, M. Baklanov and M. Popovici for sample preparation and D. Sherman  and L. Degiorgi for fruitful discussion. The work of TB was supported by the Program ``Quantum Mesoscopic and Disordered Systems" of the Russian Academy of Sciences and by the Russian Foundation for Basic Research (Grant No. 12-02-00152).

\appendix*
\section{THz properties of the Si substrate }
Since our measurements of TiN thin films were carried out as combined measurements of film and substrate, we had to analyze the properties of the bare substrate in order to  disentagle TiN film and Si substrate contributions. We therefore measured a $1\times 1$cm$^2$ slab (thickness 0.78 mm) of the same type of doped silicon, onto which our films are deposited. We found our raw data to be in excellent agreement with previous THz-time-domain measurements\cite{nas01} of doped silicon, and thus used the same way of analysis. In Fig. 5,  $T_s$ versus temperature and frequency is displayed. Starting at room temperature, $T_s$ is about 0.4, slightly decreasing towards lower frequencies. Upon cooling, $T_s$   is reduced and decreases to zero at low frequencies at around 100 K. $T_s$ rises again upon further cooling, and reaches unity at 50 K for each frequency. We do not observe changes of the Fabry-Perot oscillations  between 50-2 K, which indicates a purely dielectric Si substrate. 

Fig. 6c displays $\sigma_1$ versus temperature and frequency. For a given $T\geq 50$ K, $\sigma_1$ decreases monotonically towards higher frequencies. The slope is maximal at around 100 K. From 100-50 K, the dispersion dies off rapidly and we do not observe any residual conductivity for lower temperatures. Our single-peak fitting routine does not assume any particular electronic model, i.e. the extracted complex conductivity can be utilzed to identify the most suited theory describing the charge carrier dynamics of our substrate. Fig.7 shows $\sigma_1$ and $\sigma_2$ versus frequency for three different temperatures together with fits according to the Drude model\cite{dre02}
\begin{figure}
\includegraphics[width=8cm]{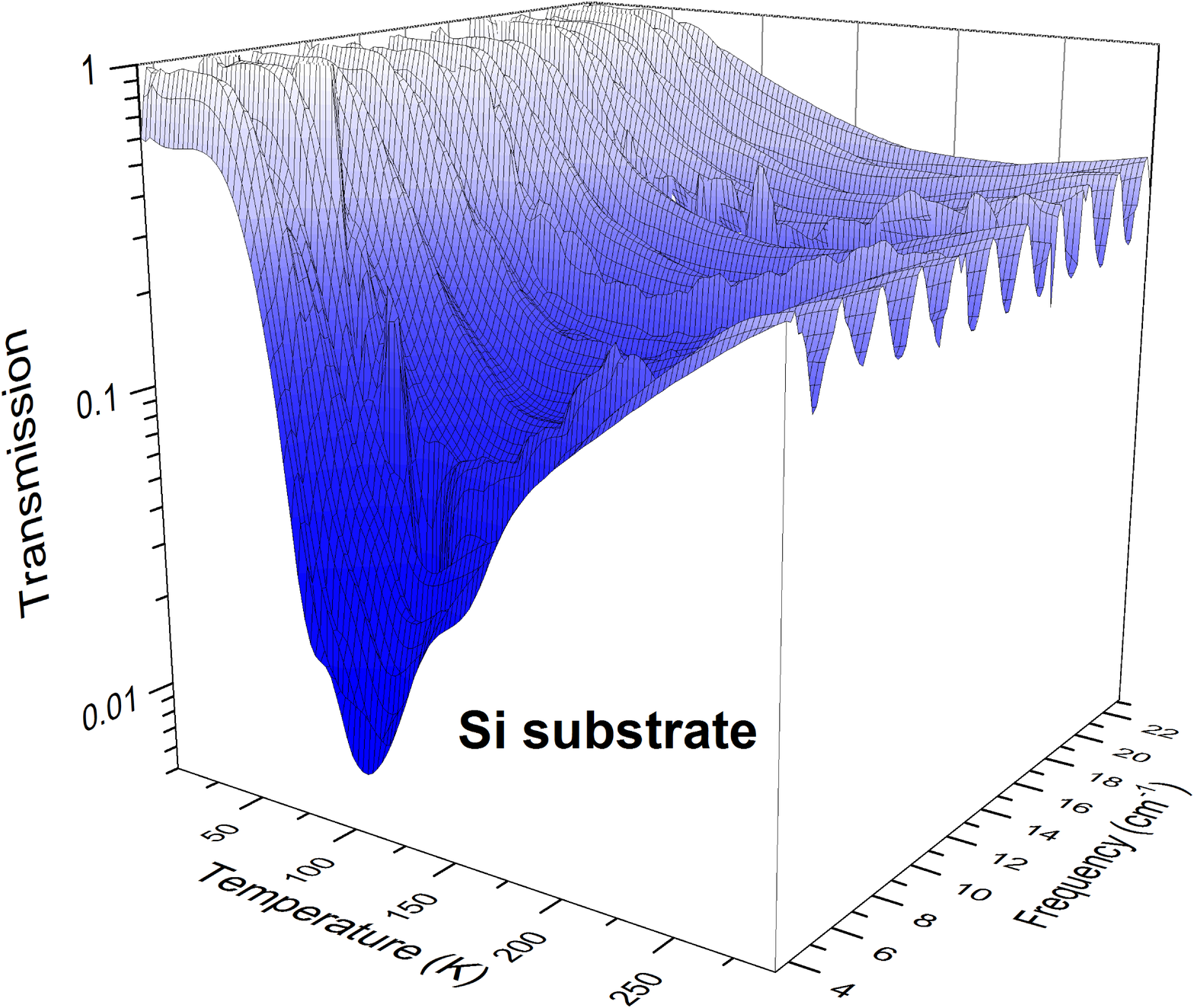}
\caption{\label{app1}(color online) $T_s$ of silicon substrate versus temperature and frequency. The frequency dependence shows the well developed Fabry-Perot oscillations, that extend over the entire temperature and spectral range except for the low-$T_s$ region at low frequencies around 100 K.}
\end{figure}
\begin{figure}
\includegraphics[width=8.66cm]{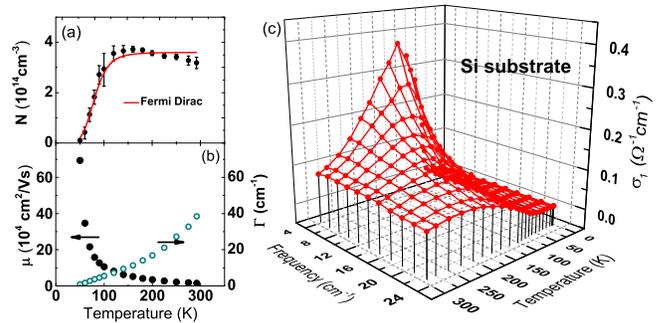}
\caption{\label{app2}(color online) Charge carrier dynamics extracted from $T_s$. (a) Charge carrier concentration (solid line: Fermi-Dirac distribution) (b) mobility and scattering rate are determined by fitting $\sigma_1$ within the Drude model. (c) real part of conductivity $\sigma_1$ versus frequency and temperature. The data is extracted from $T_s$ by single-peak fitting.}
\end{figure}
\begin{figure}
\includegraphics[width=8.66cm]{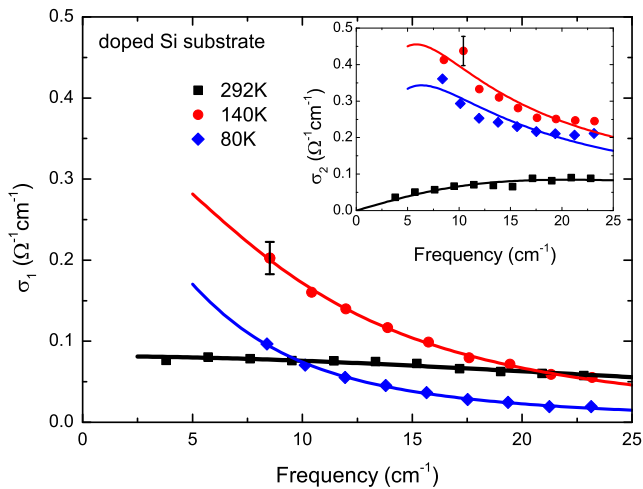}
\caption{\label{app3} (color online) Real part $\sigma_1$ and imaginary part $\sigma_2$ (inset) of the complex conductivity versus frequency for three temperatures. Solid lines are fits according to eqs. A.1 \& A.2. }
\end{figure}
\begin{equation}
\sigma_1(\omega,T)+i\sigma_2(\omega,T)=\frac{\sigma_0(T) }{1-\omega^2 \Gamma(T)^{-2}}+i\frac{\sigma_0(T) \omega \Gamma(T)^{-1}}{1-\omega^2 \Gamma(T)^{-2}} \\    
\end{equation}
where $\sigma_0=N\Gamma m e^{-2}$ is the dc conductivity\cite{dre02} ($e$ elementary charge, $m$ electron mass) and $\Gamma=\tau^{-1}$ is the scattering rate (inverse scattering time), both used as fitting parameters. Clearly, the Drude model is well suited to describe the charge carrier dynamics, however, small deviations of $\sigma_2$ are apparent. We assume that the deviations are caused by $T_s$ being close to zero and therefore not showing pronounced Fabry-Perot oscillations. This does not affect $\sigma_1$, since this quantity is related to the amplitude of $T_s$. However, it affects $\sigma_2$, since this quantity depends on the position of the oscillations with respect to the frequency, which can not be assigned with sufficient accuracy here.  
Utilizing the Drude fits of $\sigma_1$, we calculated the charge carrier concentration $N$ shown in Fig.6a. The carrier concentration $N$ is supposed to satisfy the Fermi-Dirac distribution that reads\cite{nas01}
\begin{equation}
\frac{N^{2}}{N_{D}-N}=N_{v}\left(\frac{2\pi m_{n}k_{B}T}{h^{2}}\right)^{\frac{3}{2}}\exp\left(-\frac{\Delta E_{D}}{k_{B}T}\right)
\end{equation}
with $N_D$ the donor density, $N_v$=6 the number of equivalent valleys for Si, $m_n=0.32m_0$ the effective mass, $h$ Planck constant, $k_B$ Boltzmann constant and $\Delta E_D$ the donor ionization energy. The solid line in Fig. 6a is the Fermi-Dirac distribution with $\Delta E_D=62$ meV and $N_D=3.5\times 10^{14}$ $\text{cm}^{-3}$. Localization of the charge carriers explains the vanishing conductivity at low temperatures as soon as a certain activation temperature is gone below.  Fig. 6b shows the mobility $\mu$ and the scattering rate\cite{dre02} $\Gamma = e(m\mu)^{-1}$ extracted from the Drude fits. With increasing temperature, the charge carrier movement is constrained more and more due to the enhanced phonon scattering, leading to a reduced mobility and increasing scattering rate. 

In conclusion, our Si substrate shows charge carrier dynamics, which can be assigned to the doping. Above an activation temperature of about 50 K, the frequency dependence of the complex conductivity is well fitted by the Drude model. Below 50 K, the Si substrate is purely dielectric, which enables us to unambigously disentangle Si substrate and TiN film contributions.

\bibliography{baturina10.bib}

\begin{thebibliography}{24}%
\makeatletter
\providecommand \@ifxundefined [1]{%
 \@ifx{#1\undefined}
}%
\providecommand \@ifnum [1]{%
 \ifnum #1\expandafter \@firstoftwo
 \else \expandafter \@secondoftwo
 \fi
}%
\providecommand \@ifx [1]{%
 \ifx #1\expandafter \@firstoftwo
 \else \expandafter \@secondoftwo
 \fi
}%
\providecommand \natexlab [1]{#1}%
\providecommand \enquote  [1]{``#1''}%
\providecommand \bibnamefont  [1]{#1}%
\providecommand \bibfnamefont [1]{#1}%
\providecommand \citenamefont [1]{#1}%
\providecommand \href@noop [0]{\@secondoftwo}%
\providecommand \href [0]{\begingroup \@sanitize@url \@href}%
\providecommand \@href[1]{\@@startlink{#1}\@@href}%
\providecommand \@@href[1]{\endgroup#1\@@endlink}%
\providecommand \@sanitize@url [0]{\catcode `\\12\catcode `\$12\catcode
  `\&12\catcode `\#12\catcode `\^12\catcode `\_12\catcode `\%12\relax}%
\providecommand \@@startlink[1]{}%
\providecommand \@@endlink[0]{}%
\providecommand \url  [0]{\begingroup\@sanitize@url \@url }%
\providecommand \@url [1]{\endgroup\@href {#1}{\urlprefix }}%
\providecommand \urlprefix  [0]{URL }%
\providecommand \Eprint [0]{\href }%
\providecommand \doibase [0]{http://dx.doi.org/}%
\providecommand \selectlanguage [0]{\@gobble}%
\providecommand \bibinfo  [0]{\@secondoftwo}%
\providecommand \bibfield  [0]{\@secondoftwo}%
\providecommand \translation [1]{[#1]}%
\providecommand \BibitemOpen [0]{}%
\providecommand \bibitemStop [0]{}%
\providecommand \bibitemNoStop [0]{.\EOS\space}%
\providecommand \EOS [0]{\spacefactor3000\relax}%
\providecommand \BibitemShut  [1]{\csname bibitem#1\endcsname}%
\let\auto@bib@innerbib\@empty
\bibitem [{\citenamefont {Gantmakher}\ and\ \citenamefont
  {Dolgopolov}(2010)}]{Gan10}%
  \BibitemOpen
  \bibfield  {author} {\bibinfo {author} {\bibfnamefont {V.~F.}\ \bibnamefont
  {Gantmakher}}\ and\ \bibinfo {author} {\bibfnamefont {V.~T.}\ \bibnamefont
  {Dolgopolov}},\ }\href@noop {} {\bibfield  {journal} {\bibinfo  {journal}
  {Physics-Uspekhi}\ }\textbf {\bibinfo {volume} {53}},\ \bibinfo {pages} {1}
  (\bibinfo {year} {2010})}\BibitemShut {NoStop}%
\bibitem [{\citenamefont {Vissers}\ \emph {et~al.}(2012)\citenamefont
  {Vissers}, \citenamefont {Weides}, \citenamefont {Kline}, \citenamefont
  {Sandberg},\ and\ \citenamefont {Pappas}}]{vis12}%
  \BibitemOpen
  \bibfield  {author} {\bibinfo {author} {\bibfnamefont {M.~R.}\ \bibnamefont
  {Vissers}}, \bibinfo {author} {\bibfnamefont {M.~P.}\ \bibnamefont {Weides}},
  \bibinfo {author} {\bibfnamefont {J.~S.}\ \bibnamefont {Kline}}, \bibinfo
  {author} {\bibfnamefont {M.}~\bibnamefont {Sandberg}}, \ and\ \bibinfo
  {author} {\bibfnamefont {D.~P.}\ \bibnamefont {Pappas}},\ }\href@noop {}
  {\bibfield  {journal} {\bibinfo  {journal} {Applied Physics Letters}\
  }\textbf {\bibinfo {volume} {101}},\ \bibinfo {pages} {022601} (\bibinfo
  {year} {2012})}\BibitemShut {NoStop}%
\bibitem [{\citenamefont {Sac\'ep\'e}\ \emph {et~al.}(2008)\citenamefont
  {Sac\'ep\'e}, \citenamefont {Chapelier}, \citenamefont {Baturina},
  \citenamefont {Vinokur}, \citenamefont {Baklanov},\ and\ \citenamefont
  {Sanquer}}]{sac08}%
  \BibitemOpen
  \bibfield  {author} {\bibinfo {author} {\bibfnamefont {B.}~\bibnamefont
  {Sac\'ep\'e}}, \bibinfo {author} {\bibfnamefont {C.}~\bibnamefont
  {Chapelier}}, \bibinfo {author} {\bibfnamefont {T.~I.}\ \bibnamefont
  {Baturina}}, \bibinfo {author} {\bibfnamefont {V.~M.}\ \bibnamefont
  {Vinokur}}, \bibinfo {author} {\bibfnamefont {M.~R.}\ \bibnamefont
  {Baklanov}}, \ and\ \bibinfo {author} {\bibfnamefont {M.}~\bibnamefont
  {Sanquer}},\ }\href@noop {} {\bibfield  {journal} {\bibinfo  {journal} {Phys.
  Rev. Lett.}\ }\textbf {\bibinfo {volume} {101}},\ \bibinfo {pages} {157006}
  (\bibinfo {year} {2008})}\BibitemShut {NoStop}%
\bibitem [{\citenamefont {Sac\'ep\'e}\ \emph {et~al.}(2010)\citenamefont
  {Sac\'ep\'e}, \citenamefont {Chapelier}, \citenamefont {Baturina},
  \citenamefont {Vinokur}, \citenamefont {Baklanov},\ and\ \citenamefont
  {Sanquer}}]{bat10}%
  \BibitemOpen
  \bibfield  {author} {\bibinfo {author} {\bibfnamefont {B.}~\bibnamefont
  {Sac\'ep\'e}}, \bibinfo {author} {\bibfnamefont {C.}~\bibnamefont
  {Chapelier}}, \bibinfo {author} {\bibfnamefont {T.~I.}\ \bibnamefont
  {Baturina}}, \bibinfo {author} {\bibfnamefont {V.~M.}\ \bibnamefont
  {Vinokur}}, \bibinfo {author} {\bibfnamefont {M.~R.}\ \bibnamefont
  {Baklanov}}, \ and\ \bibinfo {author} {\bibfnamefont {M.}~\bibnamefont
  {Sanquer}},\ }\href@noop {} {\bibfield  {journal} {\bibinfo  {journal}
  {Nature Communications}\ }\textbf {\bibinfo {volume} {1}},\ \bibinfo {pages}
  {140} (\bibinfo {year} {2010})}\BibitemShut {NoStop}%
\bibitem [{\citenamefont {Baturina}\ \emph
  {et~al.}(2007{\natexlab{a}})\citenamefont {Baturina}, \citenamefont
  {Bilu\v{s}i\'{c}}, \citenamefont {Mironov}, \citenamefont {Vinokur},
  \citenamefont {Baklanov},\ and\ \citenamefont {Strunk}}]{bat07}%
  \BibitemOpen
  \bibfield  {author} {\bibinfo {author} {\bibfnamefont {T.~I.}\ \bibnamefont
  {Baturina}}, \bibinfo {author} {\bibfnamefont {A.}~\bibnamefont
  {Bilu\v{s}i\'{c}}}, \bibinfo {author} {\bibfnamefont {A.~Y.}\ \bibnamefont
  {Mironov}}, \bibinfo {author} {\bibfnamefont {V.~M.}\ \bibnamefont
  {Vinokur}}, \bibinfo {author} {\bibfnamefont {M.~R.}\ \bibnamefont
  {Baklanov}}, \ and\ \bibinfo {author} {\bibfnamefont {C.}~\bibnamefont
  {Strunk}},\ }\href@noop {} {\bibfield  {journal} {\bibinfo  {journal}
  {Physica C}\ }\textbf {\bibinfo {volume} {468}},\ \bibinfo {pages} {316}
  (\bibinfo {year} {2007}{\natexlab{a}})}\BibitemShut {NoStop}%
\bibitem [{\citenamefont {Baturina}\ \emph
  {et~al.}(2007{\natexlab{b}})\citenamefont {Baturina}, \citenamefont
  {Mironov}, \citenamefont {Vinokur}, \citenamefont {Baklanov},\ and\
  \citenamefont {Strunk}}]{bat07b}%
  \BibitemOpen
  \bibfield  {author} {\bibinfo {author} {\bibfnamefont {T.~I.}\ \bibnamefont
  {Baturina}}, \bibinfo {author} {\bibfnamefont {A.~Y.}\ \bibnamefont
  {Mironov}}, \bibinfo {author} {\bibfnamefont {V.~M.}\ \bibnamefont
  {Vinokur}}, \bibinfo {author} {\bibfnamefont {M.~R.}\ \bibnamefont
  {Baklanov}}, \ and\ \bibinfo {author} {\bibfnamefont {C.}~\bibnamefont
  {Strunk}},\ }\href@noop {} {\bibfield  {journal} {\bibinfo  {journal} {Phys.
  Rev. Lett.}\ }\textbf {\bibinfo {volume} {99}},\ \bibinfo {pages} {257003}
  (\bibinfo {year} {2007}{\natexlab{b}})}\BibitemShut {NoStop}%
\bibitem [{\citenamefont {Baturina}\ \emph {et~al.}(2005)\citenamefont
  {Baturina}, \citenamefont {Bentner}, \citenamefont {Strunk}, \citenamefont
  {Baklanov},\ and\ \citenamefont {Satta}}]{bat05}%
  \BibitemOpen
  \bibfield  {author} {\bibinfo {author} {\bibfnamefont {T.~I.}\ \bibnamefont
  {Baturina}}, \bibinfo {author} {\bibfnamefont {J.}~\bibnamefont {Bentner}},
  \bibinfo {author} {\bibfnamefont {C.}~\bibnamefont {Strunk}}, \bibinfo
  {author} {\bibfnamefont {M.~R.}\ \bibnamefont {Baklanov}}, \ and\ \bibinfo
  {author} {\bibfnamefont {A.}~\bibnamefont {Satta}},\ }\href@noop {}
  {\bibfield  {journal} {\bibinfo  {journal} {Physica B}\ }\textbf {\bibinfo
  {volume} {359}},\ \bibinfo {pages} {500} (\bibinfo {year}
  {2005})}\BibitemShut {NoStop}%
\bibitem [{\citenamefont {Feigel'man}\ \emph {et~al.}(2010)\citenamefont
  {Feigel'man}, \citenamefont {Ioffe}, \citenamefont {Kravtsov},\ and\
  \citenamefont {Cuevas}}]{fei10}%
  \BibitemOpen
  \bibfield  {author} {\bibinfo {author} {\bibfnamefont {M.~V.}\ \bibnamefont
  {Feigel'man}}, \bibinfo {author} {\bibfnamefont {L.~B.}\ \bibnamefont
  {Ioffe}}, \bibinfo {author} {\bibfnamefont {V.~E.}\ \bibnamefont {Kravtsov}},
  \ and\ \bibinfo {author} {\bibfnamefont {E.}~\bibnamefont {Cuevas}},\
  }\href@noop {} {\bibfield  {journal} {\bibinfo  {journal} {Annals of
  Physics}\ }\textbf {\bibinfo {volume} {325}},\ \bibinfo {pages} {1390}
  (\bibinfo {year} {2010})}\BibitemShut {NoStop}%
\bibitem [{\citenamefont {Dubi}\ \emph {et~al.}(2006)\citenamefont {Dubi},
  \citenamefont {Meir},\ and\ \citenamefont {Avishai}}]{dub05}%
  \BibitemOpen
  \bibfield  {author} {\bibinfo {author} {\bibfnamefont {Y.}~\bibnamefont
  {Dubi}}, \bibinfo {author} {\bibfnamefont {Y.}~\bibnamefont {Meir}}, \ and\
  \bibinfo {author} {\bibfnamefont {Y.}~\bibnamefont {Avishai}},\ }\href@noop
  {} {\bibfield  {journal} {\bibinfo  {journal} {Phys. Rev. B.}\ }\textbf
  {\bibinfo {volume} {73}},\ \bibinfo {pages} {054509} (\bibinfo {year}
  {2006})}\BibitemShut {NoStop}%
\bibitem [{\citenamefont {Bouadim}\ \emph {et~al.}(2011)\citenamefont
  {Bouadim}, \citenamefont {Loh}, \citenamefont {Randeria},\ and\ \citenamefont
  {Trivedi}}]{bou11}%
  \BibitemOpen
  \bibfield  {author} {\bibinfo {author} {\bibfnamefont {K.}~\bibnamefont
  {Bouadim}}, \bibinfo {author} {\bibfnamefont {Y.~L.}\ \bibnamefont {Loh}},
  \bibinfo {author} {\bibfnamefont {M.}~\bibnamefont {Randeria}}, \ and\
  \bibinfo {author} {\bibfnamefont {N.}~\bibnamefont {Trivedi}},\ }\href@noop
  {} {\bibfield  {journal} {\bibinfo  {journal} {Nature Physics}\ }\textbf
  {\bibinfo {volume} {7}},\ \bibinfo {pages} {884–} (\bibinfo {year}
  {2011})}\BibitemShut {NoStop}%
\bibitem [{\citenamefont {Pfuner}\ \emph {et~al.}(2009)\citenamefont {Pfuner},
  \citenamefont {Degiorgi}, \citenamefont {Baturina}, \citenamefont {Vinokur},\
  and\ \citenamefont {Baklanov}}]{pfu09}%
  \BibitemOpen
  \bibfield  {author} {\bibinfo {author} {\bibfnamefont {F.}~\bibnamefont
  {Pfuner}}, \bibinfo {author} {\bibfnamefont {L.}~\bibnamefont {Degiorgi}},
  \bibinfo {author} {\bibfnamefont {T.~I.}\ \bibnamefont {Baturina}}, \bibinfo
  {author} {\bibfnamefont {V.~M.}\ \bibnamefont {Vinokur}}, \ and\ \bibinfo
  {author} {\bibfnamefont {M.~R.}\ \bibnamefont {Baklanov}},\ }\href@noop {}
  {\bibfield  {journal} {\bibinfo  {journal} {New Journal of Physics}\ }\textbf
  {\bibinfo {volume} {11}},\ \bibinfo {pages} {113017} (\bibinfo {year}
  {2009})}\BibitemShut {NoStop}%
\bibitem [{\citenamefont {Liu}\ \emph {et~al.}(2011)\citenamefont {Liu},
  \citenamefont {Kim}, \citenamefont {Sambandamurthy},\ and\ \citenamefont
  {Armitage}}]{Arm11}%
  \BibitemOpen
  \bibfield  {author} {\bibinfo {author} {\bibfnamefont {W.}~\bibnamefont
  {Liu}}, \bibinfo {author} {\bibfnamefont {M.}~\bibnamefont {Kim}}, \bibinfo
  {author} {\bibfnamefont {G.}~\bibnamefont {Sambandamurthy}}, \ and\ \bibinfo
  {author} {\bibfnamefont {N.~P.}\ \bibnamefont {Armitage}},\ }\href@noop {}
  {\bibfield  {journal} {\bibinfo  {journal} {Phys. Rev. B.}\ }\textbf
  {\bibinfo {volume} {84}},\ \bibinfo {pages} {024511} (\bibinfo {year}
  {2011})}\BibitemShut {NoStop}%
\bibitem [{\citenamefont {Van~der Pauw}(1958)}]{vdp58}%
  \BibitemOpen
  \bibfield  {author} {\bibinfo {author} {\bibfnamefont {L.~J.}\ \bibnamefont
  {Van~der Pauw}},\ }\href@noop {} {\bibfield  {journal} {\bibinfo  {journal}
  {Philips Research Reports}\ }\textbf {\bibinfo {volume} {13}},\ \bibinfo
  {pages} {1} (\bibinfo {year} {1958})}\BibitemShut {NoStop}%
\bibitem [{\citenamefont {Baturina}\ \emph {et~al.}(2012)\citenamefont
  {Baturina}, \citenamefont {Postolova}, \citenamefont {Mironov}, \citenamefont
  {Glatz}, \citenamefont {Baklanov},\ and\ \citenamefont {Vinokur}}]{bat12}%
  \BibitemOpen
  \bibfield  {author} {\bibinfo {author} {\bibfnamefont {T.~I.}\ \bibnamefont
  {Baturina}}, \bibinfo {author} {\bibfnamefont {S.~V.}\ \bibnamefont
  {Postolova}}, \bibinfo {author} {\bibfnamefont {A.~Y.}\ \bibnamefont
  {Mironov}}, \bibinfo {author} {\bibfnamefont {A.}~\bibnamefont {Glatz}},
  \bibinfo {author} {\bibfnamefont {M.~R.}\ \bibnamefont {Baklanov}}, \ and\
  \bibinfo {author} {\bibfnamefont {V.~M.}\ \bibnamefont {Vinokur}},\
  }\href@noop {} {\bibfield  {journal} {\bibinfo  {journal} {EPL}\ }\textbf
  {\bibinfo {volume} {97}},\ \bibinfo {pages} {17012} (\bibinfo {year}
  {2012})}\BibitemShut {NoStop}%
\bibitem [{\citenamefont {Dressel}\ \emph {et~al.}(2008)\citenamefont
  {Dressel}, \citenamefont {Drichko}, \citenamefont {Gorshunov},\ and\
  \citenamefont {Pimenov}}]{dre08}%
  \BibitemOpen
  \bibfield  {author} {\bibinfo {author} {\bibfnamefont {M.}~\bibnamefont
  {Dressel}}, \bibinfo {author} {\bibfnamefont {N.}~\bibnamefont {Drichko}},
  \bibinfo {author} {\bibfnamefont {B.}~\bibnamefont {Gorshunov}}, \ and\
  \bibinfo {author} {\bibfnamefont {A.}~\bibnamefont {Pimenov}},\ }\href@noop
  {} {\bibfield  {journal} {\bibinfo  {journal} {IEEE JSTQE}\ }\textbf
  {\bibinfo {volume} {14}},\ \bibinfo {pages} {399} (\bibinfo {year}
  {2008})}\BibitemShut {NoStop}%
\bibitem [{\citenamefont {Ostertag}\ \emph {et~al.}(2011)\citenamefont
  {Ostertag}, \citenamefont {Scheffler}, \citenamefont {Dressel},\ and\
  \citenamefont {Jourdan}}]{Ost11}%
  \BibitemOpen
  \bibfield  {author} {\bibinfo {author} {\bibfnamefont {J.~P.}\ \bibnamefont
  {Ostertag}}, \bibinfo {author} {\bibfnamefont {M.}~\bibnamefont {Scheffler}},
  \bibinfo {author} {\bibfnamefont {M.}~\bibnamefont {Dressel}}, \ and\
  \bibinfo {author} {\bibfnamefont {M.}~\bibnamefont {Jourdan}},\ }\href@noop
  {} {\bibfield  {journal} {\bibinfo  {journal} {Phys. Rev. B.}\ }\textbf
  {\bibinfo {volume} {84}},\ \bibinfo {pages} {035132} (\bibinfo {year}
  {2011})}\BibitemShut {NoStop}%
\bibitem [{\citenamefont {Torgashev}\ \emph {et~al.}(2012)\citenamefont
  {Torgashev}, \citenamefont {Prokhorov}, \citenamefont {Komandin},
  \citenamefont {Zhukova}, \citenamefont {Anzin}, \citenamefont {Talanov},
  \citenamefont {Rabkin}, \citenamefont {Bush}, \citenamefont {Dressel},\ and\
  \citenamefont {Gorshunov}}]{tor12}%
  \BibitemOpen
  \bibfield  {author} {\bibinfo {author} {\bibfnamefont {V.~I.}\ \bibnamefont
  {Torgashev}}, \bibinfo {author} {\bibfnamefont {A.~S.}\ \bibnamefont
  {Prokhorov}}, \bibinfo {author} {\bibfnamefont {G.~A.}\ \bibnamefont
  {Komandin}}, \bibinfo {author} {\bibfnamefont {E.~S.}\ \bibnamefont
  {Zhukova}}, \bibinfo {author} {\bibfnamefont {V.~B.}\ \bibnamefont {Anzin}},
  \bibinfo {author} {\bibfnamefont {V.~M.}\ \bibnamefont {Talanov}}, \bibinfo
  {author} {\bibfnamefont {L.}~\bibnamefont {Rabkin}}, \bibinfo {author}
  {\bibfnamefont {A.~A.}\ \bibnamefont {Bush}}, \bibinfo {author}
  {\bibfnamefont {M.}~\bibnamefont {Dressel}}, \ and\ \bibinfo {author}
  {\bibfnamefont {B.}~\bibnamefont {Gorshunov}},\ }\href@noop {} {\bibfield
  {journal} {\bibinfo  {journal} {Physics of the Solid State}\ }\textbf
  {\bibinfo {volume} {54}},\ \bibinfo {pages} {350} (\bibinfo {year}
  {2012})}\BibitemShut {NoStop}%
\bibitem [{\citenamefont {Dressel}\ and\ \citenamefont
  {Gruener}(2002)}]{dre02}%
  \BibitemOpen
  \bibfield  {author} {\bibinfo {author} {\bibfnamefont {M.}~\bibnamefont
  {Dressel}}\ and\ \bibinfo {author} {\bibfnamefont {G.}~\bibnamefont
  {Gruener}},\ }\href@noop {} {\emph {\bibinfo {title} {Electrodynamics of
  Solids}}}\ (\bibinfo  {publisher} {Cambridge University Press},\ \bibinfo
  {year} {2002})\BibitemShut {NoStop}%
\bibitem [{\citenamefont {Kittel}(1995)}]{Kit95}%
  \BibitemOpen
  \bibfield  {author} {\bibinfo {author} {\bibfnamefont {C.}~\bibnamefont
  {Kittel}},\ }\href@noop {} {\emph {\bibinfo {title} {Introduction to Solid
  State Physics}}}\ (\bibinfo {year} {1995})\BibitemShut {NoStop}%
\bibitem [{\citenamefont {Mattis}\ and\ \citenamefont {Bardeen}(1958)}]{mat58}%
  \BibitemOpen
  \bibfield  {author} {\bibinfo {author} {\bibfnamefont {D.~C.}\ \bibnamefont
  {Mattis}}\ and\ \bibinfo {author} {\bibfnamefont {J.}~\bibnamefont
  {Bardeen}},\ }\href@noop {} {\bibfield  {journal} {\bibinfo  {journal} {Phys.
  Rev. Lett.}\ }\textbf {\bibinfo {volume} {111}},\ \bibinfo {pages} {412}
  (\bibinfo {year} {1958})}\BibitemShut {NoStop}%
\bibitem [{\citenamefont {Zimmermann}\ \emph {et~al.}(1991)\citenamefont
  {Zimmermann}, \citenamefont {Brandt}, \citenamefont {Bauer}, \citenamefont
  {Seider},\ and\ \citenamefont {Genzel}}]{Zim91}%
  \BibitemOpen
  \bibfield  {author} {\bibinfo {author} {\bibfnamefont {W.}~\bibnamefont
  {Zimmermann}}, \bibinfo {author} {\bibfnamefont {E.}~\bibnamefont {Brandt}},
  \bibinfo {author} {\bibfnamefont {M.}~\bibnamefont {Bauer}}, \bibinfo
  {author} {\bibfnamefont {E.}~\bibnamefont {Seider}}, \ and\ \bibinfo {author}
  {\bibfnamefont {L.}~\bibnamefont {Genzel}},\ }\href {\doibase
  10.1016/0921-4534(91)90771-P} {\bibfield  {journal} {\bibinfo  {journal}
  {Physica C: Superconductivity}\ }\textbf {\bibinfo {volume} {183}},\ \bibinfo
  {pages} {99 } (\bibinfo {year} {1991})}\BibitemShut {NoStop}%
\bibitem [{\citenamefont {Rickayzen}(1965)}]{Ric65}%
  \BibitemOpen
  \bibfield  {author} {\bibinfo {author} {\bibfnamefont {G.}~\bibnamefont
  {Rickayzen}},\ }\href@noop {} {\emph {\bibinfo {title} {Theory of
  Superconductivity}}}\ (\bibinfo  {publisher} {Interscience Publishers},\
  \bibinfo {year} {1965})\BibitemShut {NoStop}%
\bibitem [{\citenamefont {Tinkham}(1980)}]{tin80}%
  \BibitemOpen
  \bibfield  {author} {\bibinfo {author} {\bibfnamefont {M.}~\bibnamefont
  {Tinkham}},\ }\href@noop {} {\emph {\bibinfo {title} {Introduction to
  Superconductivity}}},\ edited by\ \bibinfo {editor} {\bibfnamefont
  {B.}~\bibnamefont {Bayne}}\ and\ \bibinfo {editor} {\bibfnamefont
  {M.}~\bibnamefont {Gardner}}\ (\bibinfo  {publisher} {McGraw-Hill, New
  York},\ \bibinfo {year} {1980})\BibitemShut {NoStop}%
\bibitem [{\citenamefont {Nashima}\ \emph {et~al.}(2001)\citenamefont
  {Nashima}, \citenamefont {Morikawa}, \citenamefont {Takata},\ and\
  \citenamefont {Hangyo}}]{nas01}%
  \BibitemOpen
  \bibfield  {author} {\bibinfo {author} {\bibfnamefont {S.}~\bibnamefont
  {Nashima}}, \bibinfo {author} {\bibfnamefont {O.}~\bibnamefont {Morikawa}},
  \bibinfo {author} {\bibfnamefont {K.}~\bibnamefont {Takata}}, \ and\ \bibinfo
  {author} {\bibfnamefont {M.}~\bibnamefont {Hangyo}},\ }\href@noop {}
  {\bibfield  {journal} {\bibinfo  {journal} {Journal of Applied Physics}\
  }\textbf {\bibinfo {volume} {90}},\ \bibinfo {pages} {837} (\bibinfo {year}
  {2001})}\BibitemShut {NoStop}%
\end{thebibliography}%

\end{document}